# Facilitating Holistic Evaluations with LLMs
## Insights from Scenario-Based Experiments

**Toru Ishida[a*], Tongxi Liu[b,] Hailong Wang[c] & William K. Cheung[a]**
[a]*Department of Computer Science, Hong Kong Baptist University*
[b]*Department of Educational Studies, Hong Kong Baptist University*
[c]*Department of Civil Engineering, The University of Tokyo*
*toru.ishida@gmail.com

**Abstract:** Workshop courses designed to foster creativity are gaining popularity. However, even experienced faculty teams find it challenging to realize a holistic evaluation that accommodates diverse perspectives. Adequate deliberation is essential to integrate varied assessments, but faculty often lack the time for such exchanges. Deriving an average score without discussion undermines the purpose of a holistic evaluation. Therefore, this paper explores the use of a Large Language Model (LLM) as a facilitator to integrate diverse faculty assessments. Scenario-based experiments were conducted to determine if the LLM could integrate diverse evaluations and explain the underlying pedagogical theories to faculty. The results were noteworthy, showing that the LLM can effectively facilitate faculty discussions. Additionally, the LLM demonstrated the capability to create evaluation criteria by generalizing a single scenario-based experiment, leveraging its already acquired pedagogical domain knowledge.

## 1. Introduction

Many higher education institutions have adopted workshop-style courses. In such courses, a faculty team is recommended to perform a holistic evaluation of students from diverse perspectives. However, faculty members often lack the time to discuss the integration of differing assessment outcomes. Evaluating essays submitted by students is particularly challenging for the faculty team. In many cases, faculty evaluations vary significantly, but the scores are simply averaged without discussion.

To overcome the above problem, this paper utilizes an LLM to facilitate an efficient integration of diverse evaluations. Note that this paper does not involve essay evaluation using LLMs, such as that by Ishida et al. (2024), but is intended solely to integrate the diverse opinions of evaluators. We conducted an experiment inspired by a case that took place during a workshop course called SHIP (Social and Human Innovation by Practical Science and Engineering) at Waseda University in 2022, where the first and third authors were working for the course. In that course, faculty members expressed differing views from various perspectives, and integrating them was not straightforward. Therefore, we designed a scenario-based experiment (Schwartz, 1996) stimulated by this real experience, intentionally incorporating typical situations in holistic evaluations.

The experiment was conducted using ChatGPT-4 on October 3, 2023. The facilitation by the LLM was commendable, even surprising education researchers with its adequacy. Furthermore, the background theories presented by the LLM provided significant learning opportunities for faculty members who are not specialists in education studies. More than that, the LLM demonstrated the ability to generalize the experiment from a single scenario and create general evaluation criteria. This capability surprised artificial intelligence researchers, as it represented a function known as Explanation-Based Learning (EBL) (Mitchell, et al., 1986), which presupposes sufficient pedagogical domain knowledge.

## 2. Research Questions and Methods

The research questions related to the facilitation of essay evaluation by LLMs are as follows:

- Can LLMs integrate diverse opinions and compile evaluation results?

    Instead of merely averaging scores, it is essential to design the process of integrating different perspectives and compiling evaluations. Additionally, it is crucial to discern which opinions should be considered and which should not, from the standpoint of fairness in evaluation.

- Can LLMs theoretically explain the background of their judgments?

    In the field of education studies, numerous theories have been accumulated. These theories are not always consciously considered in the actual essay evaluation. However, demonstrating the theoretical background when integrating different evaluations enhances persuasiveness towards faculty members. It also serves as a learning opportunity for faculty members who are not specialists in education studies.

- Can LLMs generalize an experience from a specific case and generate evaluation criteria?

    Evaluation criteria (rubrics in this context) are usually created before starting a course, alongside the syllabus, and need to be improved through experience. However, faculty often are so constrained by time that they can only manage to grade, without improving the rubric. Generalizing a specific case to create a rubric could greatly aid in improving the course.

To investigate these issues, we will conduct a scenario-based experiment (Schwartz, 1996). However, in this experiment, the subjects are not humans but LLMs. The scenarios will embed typical situations that arise during the facilitation process, specifically: 1) Compromising different opinions, 2) Evaluating students' growth, 3) Handling peer evaluations, and 4) Taking into account unique contributions.

These situations will be given to the LLMs as prompts. However, the prompts will not include any essay evaluation criteria or educational theories and background knowledge. In other words, this experiment aims to investigate the facilitation capabilities of untuned LLMs in essay evaluation. At the end of the experiment, based on the experience with the scenarios, the LLMs will be asked to create evaluation criteria.

## 3. Scenario-Based Experiments

### 3.1 Compromising Different Opinions

This experiment was conducted under the assumption that the LLM operates solely based on its knowledge and capabilities, without prompts such as the outlines of workshop courses or evaluation criteria. The scenario begins with several faculty members presenting different evaluation perspectives—focusing on motivation, technical skills, and report formatting, respectively.

**Scenario Part 1** (LLM Prompt)**:**

Three faculty members (hereinafter referred to as F1, F2, F3, etc.) wrote about their evaluations of the reflection essay of a student who participated in a technology-oriented workshop course.

**F1:** The essay shows that this student was very motivated and worked hard. Technically, I'm not sure if he understood it well, but I'd like to give him an A based on his strong motivation.

**F2:** This student expresses his enthusiasm in writing, but it is not clearly written how he actually contributed to the team. Also, I don't think he understood the technology very well. Therefore, the grade should be B.

**F3:** This student wrote significantly less text than other students' essays. We told students only an upper limit for the amount of the essay, so he should not fail the exam just because the amount is too small, but if he can't write the essay in enough length, it means he wasn't working hard enough. Therefore, the grade will be C.

The LLM facilitator, as shown in Table 1, synthesizes arguments from various perspectives (hereinafter referred to as P1, P2, P3, etc.) and leads to a conclusion. Note that all outputs from the LLM will be presented within tables hereafter.

Table 1. Evaluation by the LLM Facilitator (LLM Output)

| **Evaluation Perspectives:** |
|---|
| Based on the feedback from the three faculty members (F1, F2, and F3), the evaluations seem to revolve around three main perspectives: |
| **P1. Motivation & Enthusiasm:** |
| All faculty members acknowledged the student's motivation and enthusiasm. F1 especially valued this and recommended an A, while F2 noted it but felt there was insufficient evidence of actual contribution, suggesting a B. |
| **P2. Understanding of Technology:** |
| Both F1 and F2 mentioned concerns about the student's technical understanding, while F3 did not specifically mention this aspect. |
| **P3. Length & Depth of the Essay:** |
| F3 was concerned about the essay length, equating shorter length with less effort, leading to a C recommendation. Neither F1 nor F2 commented directly on the essay's length, but F2 did mention that the student did not clearly convey their contributions. |
| **Conclusion:** |
| The student's enthusiasm is evident, but there are significant concerns regarding their understanding and the depth of their reflection essay. Taking all opinions into account, a grade of B seems to be a balanced conclusion. However, it might also be worthwhile for the facilitator to provide specific feedback to the student regarding the areas of concern to encourage improvement in the future. |

While the LLM facilitator's decision is reasonable, it may appear as merely an average, and faculty members may not be satisfied with it. Consequently, we asked the LLM to explain the theories underlying its judgment (hereinafter referred to as T1, T2, etc.) and the reasons for choosing these particular theories.

Table 2. Theoretical Basis for the Judgment (LLM Output)

| **Background Theories:** |
|---|
| **T1. Triangulation:** |
| As mentioned, the decision was based on triangulating the different perspectives provided by the three faculty members. Triangulation aids in validation by cross-verifying from multiple sources. |
| **T2. Weighted Average Decision Making:** |
| This is a common decision-making approach used when different criteria have different importance levels. If all faculty members' opinions hold equal weight, averaging the grades proposed by each of them provided a middle ground. |

> **Reason for Choosing These Theories:**
>
> - Holistic Viewpoint: The nature of triangulation means that it aims to provide a more comprehensive view of the situation. In the context of evaluation, having multiple viewpoints can yield a more balanced and fair judgment.
> - Reduction of Bias: By considering multiple perspectives, it helps mitigate individual biases. If only one faculty member's opinion was used, the grade might be skewed based on that personal grading philosophy or emphasis on certain criteria over others.
> - Practicality: Given the information provided, triangulation offered a straightforward approach to synthesize the data into a cohesive recommendation without needing complex algorithms or additional data.

The LLM clarified that it based its decision on the principle of Triangulation (Patton, 1999). This approach employs various methods or sources of data to provide a comprehensive assessment of phenomena.

### 3.2 Evaluating Student's Growth

In the evaluation of students, a recurring debate centers on whether greater emphasis should be placed on assessing achievement or growth. Conventional lecture-based courses mainly assess achievement through final examinations. Similarly, courses that focus on skill acquisition often use rubrics to evaluate achievement. However, in workshop courses designed to foster creativity, there is a prevalent opinion that emphasizes the importance of assessing growth.

**Scenario Part 2** (LLM Prompt):

> A faculty member, F4, submitted their evaluation late. The faculty, having observed the workshop, was impressed by the significant growth demonstrated by the student.
>
> **F4:** I have been observing this student's activities in the workshop, and although he was initially self-centered, he gradually became able to collaborate. I would like to evaluate that progress. I can't say he's an A in terms of what he achieved, but if we're evaluating growth, he's clearly an A.

F4 evaluates growth as significant and recommends an A for growth, but not for his achievement. This opinion differs from F2 regarding their perspectives on the student's growth, so the LLM must lead to a balanced conclusion.

Table 3. Evaluation by the LLM Facilitator (LLM Output)

> **Evaluation Perspectives:**
>
> F4 introduced a new perspective—the student's personal growth and adaptability.
>
> **P4. Personal Growth:**
>
> > The journey from being self-centered to becoming collaborative indicates significant personal development.
>
> **Conclusion:**
>
> Taking into account all faculty opinions, and giving due weight to personal growth as highlighted by F4, a grade of B+ seems appropriate. The student has shown considerable motivation and personal growth, which is commendable. However, concerns regarding technical understanding and clarity in the reflection essay prevent a higher grade. Again, specific feedback regarding areas of improvement would be beneficial for the student's continued development.

The conclusions drawn by the LLM not only demonstrate a balanced evaluation of achievement and growth but also highlight the importance of feedback from the evaluation team. It can be said that the LLM exhibited excellent judgement and provided a valuable suggestion to faculty members. When inquiring into the underlying theories, it referenced Holistic Assessment (Wiggins, 1993) and Developmental Evaluation (Patton, 2010).

Table 4. Theoretical Basis for the Judgment (LLM Output)

| |
|---|
| **Background Theories:** |
| **T3. Holistic Assessment:** |
| This theory emphasizes considering all aspects of a learner's performance rather than focusing solely on academic achievement. Holistic assessment values non-academic attributes, such as motivation, enthusiasm, and personal growth. |
| **T4. Developmental Evaluation:** |
| This approach focuses on assessing growth and development over a specified period. Instead of evaluating the end result or outcome, developmental evaluation gives importance to the journey, recognizing and valuing the learning process. |
| **Reason for Choosing These Theories:** |
| • Comprehensive Understanding: Both theories emphasize an understanding of an individual. It ensures that we aren't just quantifying learning through grades but understanding the student's journey, personal growth, and non-academic attributes. |
| • Emphasis on Growth: F4's feedback highlighted the student's personal growth. Adopting a developmental evaluation lens ensures that this aspect isn't overlooked. Education is not just about achievements but also about development of the learner. |
| • Balancing Multiple Perspectives: The holistic assessment approach allows the integration of diverse opinions into a single cohesive judgment. By considering all these perspectives, we can ensure a more balanced and fair evaluation. |

### *3.3 Handling Peer Evaluations*

Another important topic of discussion in workshop assessments is the use of peer evaluation, where students participate in the assessment process. In situations where it is challenging for faculty to consistently observe, peer evaluation is often considered a means to assess the quality of students' engagement. However, peer evaluation can sometimes devolve into a popularity contest and potentially sour the team atmosphere. The following scenario addresses a situation where peer evaluation was incidentally mentioned in other students' essays.

**Scenario part 3** (LLM Prompt)**:**
This student was once given a B+. After reading other students' essays, however, it appears that several students reported this student often disrupted the team's coordination. However, not all students on the team reported this problem. When this matter came up for discussion, faculty members' opinions were divided. One faculty member pointed out the importance of student peer evaluation, and another argued that students were not asked to do such evaluations in the first place in this essay.

Table 5. Evaluation by the LLM Facilitator (LLM Output)

| |
|---|
| **Evaluation Perspectives:** |
| The situation described introduces a complex issue that touches upon various themes in the realm of education and evaluation: |

> **P4. Validity of Peer Evaluations:**
>
> Peer evaluations can provide valuable insights into team dynamics and interpersonal skills. They allow for a more holistic understanding of an individual's contribution and behavior within a group. However, their subjective nature may also introduce biases.
>
> **P5. Consistency in Assessment:**
>
> The course evidently did not ask for or mandate peer evaluations. Incorporating them after the fact can introduce inconsistency in the evaluation process.
>
> **P6. Incomplete Feedback:**
>
> Only some students provided feedback on the student's disruptive behavior. This incomplete representation may not accurately capture the entire team's experience.
>
> **Conclusion:**
>
> Considering the complexities, it might not be fair to adjust the student's grade based on feedback that was not initially part of the assessment criteria. However, it's crucial to address the concerns raised by team members. Offering the student feedback and support can ensure personal growth and better team dynamics. If peer evaluations are deemed valuable, they should be formally integrated into the assessment process in future instances, ensuring consistency and fairness in evaluations.

The LLM demonstrated its capability to handle complex issues. While acknowledging the significance of peer evaluation, concerns have been raised about the evaluation becoming inconsistent due to not all students participating. The conclusion reached is that students' opinions should not be reflected in this evaluation. This is a highly logical and valid conclusion, and the subsequent explanation, based on the theory of Constructive Alignment (Biggs, 1996) and Reliability in Assessment (Nitko and Brookhart, 2007), is outstanding.

Table 6. Theoretical Basis for the Judgment (LLM Output)

> **Background Theories:**
>
> **T5. Constructive Alignment:**
>
> This theory emphasizes that all components of a course – its learning objectives, teaching methods, and assessment tasks – should be aligned to ensure consistency and fairness. Introducing a new assessment criterion (peer evaluations) after grades have already been assigned would violate this principle.
>
> **T6. Reliability in Assessment:**
>
> Reliability pertains to the consistency and stability of assessment scores. It is crucial that an assessment approach yields similar results under consistent conditions. Introducing peer feedback after grades have already been determined can compromise the reliability of the assessment process, as the new information could introduce variability not present in the original evaluations.
>
> **Reason for Choosing These Theories:**
>
> - Consistency in Assessment: Constructive Alignment highlights the importance of maintaining consistency between the established objectives, teaching methods, and assessment criteria.
>
> - Stability of Grades: If grades can be retroactively altered based on new criteria or feedback, this can undermine trust in the grading system. Reliability in Assessment emphasizes the importance of maintaining stability and consistency in grading.
>
> - Fairness and Trust: Both theories, when applied together, ensure that the evaluation process is transparent, reliable, and aligned. This fosters trust in the system, ensuring students believe in the fairness and accuracy of their evaluations.

## 3.4 Taking into Account Unique Contributions

In cases where faculty recognize the unique talents and contributions of students, it may be proposed that additional points be awarded. This typically occurs in workshop courses designed to foster creativity. In the scenario below, an experienced faculty member advocates for extra points for a student.

**Scenario Part 4** (LLM Prompt)**:**

During the grading discussion, an experienced faculty member, F5, joined and expressed his opinion strongly. This faculty member does not have the authority to evaluate grades, but other faculty members respect him because of his experience.

**F5:** He's a genius. His words changed the direction of the discussion in that team. In workshop subjects like this, the grading criteria should be different from regular subjects where students get good grades just because they are excellent. If we don't evaluate these kinds of talents, we don't need such a course in the first place.

Table 7. Evaluation by the LLM Facilitator (LLM Output)

| |
|---|
| **Evaluation Perspectives:** |
| Given the perspectives, the grades to consider are three A's, one B, and one C. It would lean towards a grade of A-. However, given F5's experience and respect among peers, his evaluation might carry additional weight, especially when evaluating unconventional or outstanding contributions in a workshop setting. |
| **P7. Unique Contributions:** |
| F5 brings forward the idea that the student's contributions are game-changing, which could be crucial in a workshop setting. |
| **Conclusion:** |
| Given the diverse evaluations and the importance of F5's perspective, a grade of A- seems appropriate. This grade acknowledges the student's unique contributions and growth, as highlighted by F4 and F5, while also considering the reservations expressed by F1, F2, and F3. The grade recognizes the student's unique value while ensuring that fundamental course objectives are maintained. |

The conclusion of the LLM above respects the opinion of experienced faculty and awards extra points for this student's unique contributions. The underlying theories mentioned are Multiple Intelligences (Gardner, 1983) and Epistemic Authority (Brante, 2001). However, the fact that this faculty member is not part of the evaluation committee may necessitate a discussion about the fairness of the evaluation process.

Table 8. Theoretical Basis for the Judgment (LLM Output)

| |
|---|
| **Background Theories:** |
| **T7. Multiple Intelligences:** |
| This theory identified multiple distinct intelligences, suggesting that students can be proficient in one form of intelligence but not necessarily in others. The theory emphasizes that students have unique strengths that may not always fit traditional assessment metrics. |
| **T8. Epistemic Authority:** |
| This concept comes from the field of sociology of knowledge. It refers to the recognition and deference given to individuals based on their perceived knowledge, wisdom, or experience in a particular domain. F5 represents an epistemic authority due to his experience and the respect he commands among peers. |

> **Reason for Choosing These Theories:**
> - Diverse Forms of Excellence: Traditional metrics may not always capture all forms of intelligence or excellence. Recognizing these unconventional strengths, especially in a workshop setting, is crucial to fostering diverse talents.
> - Valuing Expertise: The principle of Epistemic Authority was chosen to emphasize the importance of giving due weight to the insights of those with profound expertise, especially when their insights challenge conventional wisdom or established norms.

## 3.5 Creating Evaluation Criteria

We tasked an LLM with creating evaluation criteria (rubrics in this context) based solely on the above scenario-based experiments. The prompt given to the LLM was, "Based on the above experiment, create general evaluation criteria for students' essays." No additional information related to the evaluation criteria was provided, other than the experiment scenario and the discussions. The LLM created the following evaluation criteria.

Table 9. Evaluation Criteria (LLM Output)

> **Evaluation Criteria:**
> Given the insights derived from the faculty members' perspectives and the theories discussed, here's a suggested scoring criteria for the students' essays in a technology-oriented workshop. (Total: 100 points)
> **1. Technical Understanding** (25 points)
> - Demonstrates a thorough grasp of technical concepts
> - Effectively applies technical knowledge in workshop activities
>
> **2. Motivation & Enthusiasm** (20 points)
> - Expresses a strong drive and keen interest in the workshop topics
> - Shows consistent effort and engagement throughout the course
>
> **3. Contribution to Team & Collaboration** (20 points)
> - Describes specific contributions to the team's tasks
> - Demonstrates growth in collaboration and teamwork skills
>
> **4. Unique Insights & Game-Changing Contributions** (15 points)
> - Brings forward innovative ideas or perspectives
> - Influences team's direction or sparks meaningful discussions
>
> **5. Reflective Depth** (10 points)
> - Discusses personal growth, realizations, and learning experiences
> - Demonstrates introspection about challenges and how they were overcome
>
> **6. Essay Structure & Presentation** (5 points)
> - Organizes essay in a coherent and logical manner
> - Uses clear language, proper grammar, and appropriate formatting
>
> **7. Length & Completeness** (5 points)
> - Provides adequate details within the given parameters
> - Covers all relevant aspects of the workshop experience

Of the generated evaluation criteria, the perspectives of Technical Understanding, Motivation and Enthusiasm, and Length and Completeness are associated with Scenario Part 1. Reflective Depth relates to Scenario Part 2, Contribution to Team and Collaboration to Scenario Part 3, and Unique Insights and Game-Changing Contributions to Scenario Part 4. The LLM generated evaluation criteria with such distinct relationships to the scenarios utilized in this experiment. This demonstrates the LLM's capability to generalize from one specific evaluation experiment to formulate evaluation criteria.

## 4. Insights

In the previous section, we conducted an experiment with a scenario that included 1) Compromising different opinions, 2) Evaluating students' growth, 3) Handling peer evaluations, and 4) Taking into account unique contributions. After the experiment, we tasked the LLM with creating evaluation criteria. Although not detailed in this paper due to space limitations, subsequent discussions have addressed the appropriate level of detail in rubrics for holistic evaluation and the development of scoring-free evaluation guidelines. Furthermore, the LLM has provided explanations for faculty on why and how to use the evaluation criteria. The key insights from this series of discussions with LLMs are as follows:

1. Facilitation capability:

   The experiments described in the previous section clearly demonstrate that LLMs possess significant facilitation capabilities in evaluating student essays. While education researchers should possess the requisite knowledge and ability, faculty members on each evaluation committee do not always have the ability to articulate and consolidate differing opinions to this extent. Therefore, it may be appropriate to welcome LLMs not just as assistants, but as partners in evaluation committees. However, this would necessitate careful discussion regarding the social acceptability of LLMs, including whether to disclose LLM evaluations to students.

2. Capability to present various theories and literature:

   When asked about the rationale behind its decisions, the LLM can present the underlying theories and literature. The depth of knowledge and understanding displayed by the LLM is striking. In the previous sections, we report eight theories and seven related pieces of literature, but including subsequent discussions, a total of 23 theories have been introduced by the LLM. Although it is known that LLMs sometimes suggest non-existent papers, all literature presented in this experiment was real, indicating that using LLMs as facilitators allows for learning from a multitude of theories and literature.

3. Generalization capability from a single experiment to evaluation criteria:

   The LLM has demonstrated the ability to generate evaluation criteria from a single experiment. This signifies that the LLM has achieved generalization from a single case, realizing what was proposed in the field of artificial intelligence as Explanation-Based Learning (EBL) (Mitchell, et al., 1986). EBL required logically organized domain knowledge as a prerequisite, and it is considered that the LLM has used machine-learned domain knowledge for generalization.

## 5. Conclusion

In this paper, we conducted an experiment to integrate diverse assessments into holistic evaluation, exploring the potential of LLMs as facilitators. The scenarios used in the experiments included: 1) Compromising different opinions, 2) Evaluating students' growth, 3) Handling peer evaluations, and 4) Taking into account unique contributions, highlighting challenges in essay evaluation. After resolving the cases that incorporated typical challenges, we derived general evaluation criteria from the LLM. The results revealed the LLM's: 1) Facilitation capability, 2) Capability to present various theories and literature, and 3) Generalization capability from scenario to evaluation criteria.

The experiments demonstrated that LLMs possess sufficient knowledge and facilitation capabilities to participate in essay evaluation committees. More importantly, it was shown that faculty can interact with the LLM to interpret cases and solve problems by applying relevant theories when faced with complex situations. This practical learning opportunity, previously unavailable, indicates that LLMs can be powerful partners in essay evaluation.

Finally, we must mention that integrating LLMs into educational assessments raises significant ethical considerations. For example, some subtle mistakes or biases in the LLMs' evaluations are still inevitable, which, if not detected, can cause unfair evaluations. Furthermore, issues such as fairness and transparency in the use of LLMs are paramount. Potential biases arising from different groups of essay data can lead to unjust outcomes in essay evaluations. It is also essential for educators to understand how LLMs make evaluation decisions. Developing transparent guidelines and frameworks is crucial for human-LLM collaboration in essay evaluation tasks.

## Acknowledgements

This research originated from the interdisciplinary educational experiences at Waseda University and has evolved through multidisciplinary discussions in the fields of education and computer science. We would like to express our sincere gratitude to Vivien Shuo Zhou and Benjamin Luke Moorhouse for their invaluable support throughout this research.